\documentclass[preprintnumbers,eqsecnum,aps,prd,epsf,showpacs,nofootinbib,twocolumn]{revtex4}
\usepackage{latexsym,amsmath,amssymb}
\usepackage[dvips]{graphicx,color}% Include figure files
\usepackage{dcolumn,bm}    % Align table columns on decimal point
\usepackage{subfigure}  % sub labeling of figures
\newcommand{\gsim}{\mbox{\raisebox{-1.ex}{$\stackrel
      {\textstyle>}{\textstyle\sim}$}}}
\newcommand{\lsim}{\mbox{\raisebox{-1.ex}{$\stackrel
      {\textstyle<}{\textstyle \sim}$}}}

\newcommand{\calO}{{\cal O}}
\begin{document}
\thispagestyle{empty}
\title{Application of beyond $\delta N$ formalism \\
--Varying sound speed--}

\author{Yu-ichi Takamizu}
\email{takamizu_at_yukawa.kyoto-u.ac.jp,
yt313_at_cam.ac.uk}

\affiliation{
Yukawa Institute for Theoretical Physics, Kyoto Universty, Kyoto 606-8502, Japan \\
Department of Applied Mathematics and Theoretical Physics, University of Cambridge, Wilberforce Road, Cambridge CB3 0WA, UK}
\date{\today}

\begin{abstract}
We focus on the evolution 
of curvature perturbation on superhorizon scales by adopting 
the spatial gradient expansion and show that 
the nonlinear theory, called the beyond $\delta N$-formalism as the next-leading order in the expansion. 
As one application of our formalism for a single scalar field, we investigate 
the case of varying sound speed. 
In our formalism, we can deal with the 
time evolution in contrast to $\delta N$-formalism, where 
curvature perturbations remain just constant, and nonlinear curvature perturbation follows the simple master equation whose form is similar as one in linear theory. So the calculation of bispectrum can be done in the next-leading order in the expansion as similar as the case of deriving the power spectrum. We discuss localized features of both primordial power and bispectrum 
generated by the effect of varying sound speed with a finite 
duration time. We can see a local feature like a bump in 
the equilateral bispectrum. 
\end{abstract}
\pacs{98.80.-k, 98.90.Cq}
\maketitle

\section{Introduction}
Recent observations of the cosmic microwave background anisotropy, such as WMAP and PLANCK satellites \cite{Ade:2013ktc, Bennett:2012zja} show very good agreement of the observational data with the prediction of standard inflationary cosmology where primordial fluctuations generated from quantum fluctuations of an inflaton field (see \cite{Lyth-book} for review). 
The most recent observations by the PLANCK \cite{Ade:2013uln,Ade:2013ydc} show that the primordial curvature perturbation has nearly scale invariant and follows almost perfect Gaussian statistics. The non-Gaussianity of primordial fluctuations is a powerful probe to discriminate inflationary models and also distinguish among different models (see, {\em e.g.}, Ref. \cite{CQG-focus-NG} and 
references therein). Therefore, if any tiny signature from these observations will be detected, it can tell us important information on the physics behind inflation. 
The PLANCK data \cite{Ade:2013ydc} 
have measured $-8.9<f^{loc}_{NL}<14.3$ and $-192<f^{eq}_{NL}<108$  for the so-called local type and equilateral type of non-Gaussianity, respectively, at the 2$\sigma$(95\%) confidence level. These observations show that the primordial curvature perturbation follows almost perfect Gaussian statistics, however it may 
be detected at smaller scales and also as some tiny localized feature in the bispectrum. Especially, although the quantity $f_{NL}^{loc}$ is now constrained very strongly, the possibility still remains that non-Gaussianity of equilateral shape has localized features. If ever detected, it would tell us important properties of the curvature perturbation and be a probe to distinguish the models of 
inflation. 

The gradient expansion approach \cite{Salopek:1990jq,Nambu:1994hu,
Sasaki:1998ug, Wands:2000dp, 
Rigopoulos:2003ak,Lyth:2005fi,Takamizu:2010je,Takamizu:2010xy,Takamizu:2008ra,Tanaka:2007gh,Tanaka:2006zp,Naruko:2012fe,Takamizu:2013gy} to discuss the evolution of 
nonlinear curvature 
perturbation on superhorizon scales is a powerful tool on calculation as well as the second-order perturbation theory \cite{Maldacena:2002vr,Malik:2003mv}. 
The lowest order in the expansion is the so-called $\delta N$-formalism \cite{Sasaki:1998ug,Lyth:2005fi}. 
But in order to analyzing local features of the equilateral bispectrum, 
this formula is not suitable since it leads to that nonlinearity of curvature perturbations on long-wavelength scales over horizon always generates the local shape of bispectrum. Therefore we focus on the nonlinear theory valid to the next-leading order in the expansion. It called the beyond $\delta N$-formalism and it is able to give us not only the local, but also the equilateral shape of bispectrum in contrast to $\delta N$-formalism, even though the expansion technique is taken on superhorizon scales \cite{Takamizu:2010xy,Naruko:2012fe}. 
Our nonlinear theory of the 
next-leading order in the expansion includes such subhorizon effect corresponding to the equilateral shape by matching a superhorizon curvature perturbation and subhorizon one suitably. 

The main purpose of this paper is to investigate the situation where effective sound speed changes with a finite duration time 
and analyze whether features can 
appear in the bispectrum, in particular of the equilateral shape 
by using our nonlinear perturbation theory. The previous papers have studied 
the models of varying sound speed both in the power spectrum 
and in the bispectrum \cite{Khoury:2008wj,Park:2012rh,Ribeiro:2012ar,Emery:2012sm,Achucarro:2012fd,Nakashima:2010sa,Bartolo:2013exa,Achucarro:2013cva} (See also {\em e.g.}, \cite{Achucarro:2010da,Saito:2012pd,Saito:2013aqa} for the heavy physics, related to the 
same purpose and references therein), where one basically assumed a sudden change, however, we particularly focus on the effect of a finite duration time. As a simple application of 
beyond $\delta N$-formalism, we will consider a single scalar field whose 
effective sound speed will change in time due to a non-canonical kinetic term. 
As a first step, we will assume the background evolution follows a simple slow-roll inflation, although more realistic situation would be realized for a more complicated coupled kinetic term on multi-scalar system, such as the 
curvaton scenario \cite{Moroi:2001ct,Lyth:2001nq}, 
otherwise the slow-roll conditions will be also violated. 
However in this paper in order to extract the effects of the change of 
sound speed alone, we study the case of varying sound speed without 
affecting the background evolution as a simple tractable example, as same setup as in Ref. \cite{Nakashima:2010sa} and see the appendix therein for more detailed discussion.

The rest of the paper is organized as follows. In Sec. II, we review beyond $\delta N$-formalism and especially, focus on the point that the master equations of curvature perturbation in linear and nonlinear theory show similar forms and derive the calculations. Then we discuss the possible example as the case of varying sound speed in Sec. III and derive featured power spectrums and bispectra of the curvature perturbations affected by such changes. Section IV is devoted to the conclusion.

%%%%%%%%%%%%%%%%%%%%%%%%%%%%%%%%%%%%%%%%%%%%%%%%%%%%%%%%
\section{Beyond $\delta N$-Formalism }
The gradient expansion technique
has been applied up to second order in the expansion 
to a universe dominated by a single \cite{Tanaka:2006zp,Tanaka:2007gh,
Takamizu:2008ra,Takamizu:2010xy} and multi-scalar field \cite{Naruko:2012fe}, 
yielding the formalism ``beyond $\delta N$''. 
The formulae have been also extended to be capable of 
a universe filled with a most generic non-canonical scalar field \cite{Takamizu:2013gy}, which can give the so-called G-inflation.
In this paper, we will consider a single scalar field as a simple example, whose kinetic term is a non-canonical, whose Lagrangian takes the form; $P(X,\phi)$ where $X=-\partial^{\mu}\phi\partial_{\mu}\phi/2$ because we will later discuss the situation the effective sound speed: $c_s^2=P_X/(P_X+2P_{XX}X)$ changes in time where the subscript $X$ represent derive with respect to $X$ and 
notice that the Lagrangian denoted by $P$ plays the role of the pressure as shown in Ref \cite{Takamizu:2008ra,Christopherson:2008ry}. 

Following Ref. \cite{Takamizu:2010xy}, we will briefly review beyond 
$\delta N$-formalism for a single scalar field in this section. 
This system is characterized by a single scalar degree of freedom, and hence one expects that a single master variable governs
the evolution of scalar perturbations even at nonlinear order. 
By virtue of gradient expansion, one can indeed derive a simple evolution equation
%%%%%%%%%%%%%%%%%%%%%%%%%%%%%%%%%%%%%%%%%%%%%%%%%%%%
\footnote{Also for a generic non-canonical single scalar field, the master equation becomes a simple evolution equation as a same form. As shown 
in Ref. \cite{Takamizu:2013gy}, 
the system described by the so-called G-inflation, that is $P(X,\phi)-
G(X,\phi) \Box \phi$ 
can be reduced to a same form with a extended definition of $z$.}
%%%%%%%%%%%%%%%%%%%%%%%%%%%%%%%%%%%%%%%%%%%%%%%%%%%%%%%%%%%%%%%%
for an appropriately defined master variable ${\cal R}_c^{\rm NL}$ on comoving hypersurfaces:
\begin{eqnarray}
{{\cal R}_c^{\rm NL}}''+2 {z'\over z} 
{{\cal R}_c^{\rm NL}}' +{c_s^2\over 4} \,R^{(2)}[\,
{\cal R}_c^{\rm NL}\,]=\calO(\epsilon^4)\,,\label{ieq1}
\end{eqnarray}
with 
\begin{eqnarray}
z\equiv {a\over H}\left({\rho+P\over c_s^2}\right)^{1\over 2}\,,
\label{zdef}
\end{eqnarray}
where $\rho$ and $P$ denote energy density and pressure of a scalar field, 
respectively, 
the prime represents differentiation with respect to the conformal time 
$\tau$,
$\epsilon$ is the small expansion parameter, and $R^{(2)}[X]$ is 
the Ricci scalar of the metric $X$. Hereafter we attach the superscript 
$(m)$ to a quantity  of order of gradient expansion: $\calO(\epsilon^m)$ 
and take the metric as 
\begin{align}
ds^2=a^2(-\alpha^2 d\tau^2 + e^{2\ell}\delta_{ij}dx^idx^j)+\calO(\epsilon^3)\,,
\end{align} 
where $\alpha$ denotes the lapse function, while the shift vector is vanishing at the next-leading order
$\beta^i=\calO(\epsilon^3)$,  
then $R^{(2)}[X]$ can be given by 
\begin{align}
R^{(2)}[\ell^{(0)}]=\, 
{-2 (2 \Delta \ell^{(0)}+\delta^{ij}\partial_i \ell^{(0)}
\partial_j \ell^{(0)})e^{-2 \ell^{(0)}}}\,.
\label{def: R2}
\end{align}
The equation (\ref{ieq1}) is to be compared with its linear counterpart:
\begin{eqnarray}
{{\cal R}^{\rm Lin}_c}''+2{z'\over z} {{\cal R}^{\rm Lin}_c}'
-c_s^2\,\Delta {\cal R}^{\rm Lin}_c=0\,,\label{ieq2}
\end{eqnarray}
from which one notices
the correspondence between the linear and nonlinear evolution equations. 
In order to calculate the evolution equations in 
Fourier space, we have to take the replacement $\Delta\to -k^2$. 

It is important notice that the structures  of both 
(\ref{ieq1}) and (\ref{ieq2}) are similar forms, except for last terms in the left hand sides. This point is advantage in order to estimate evolutions of curvature perturbations in linear and nonlinear theory since same calculation is 
valid  on following the evolution equation. We will see the details later.
%%%%%%%%%%%%%%%%%%%%%%%%%%%%%%%%%%%%%%%%%%%%%%%%%%%
\subsection{Linear Theory valid up to $\calO(\epsilon^2)$}
To obtain the power spectrum, we will use the linear theory of the curvature perturbation in this subsection. 
The above equation (\ref{ieq2}) has two independent solutions; conventionally 
called a growing mode and a decaying mode. We assume that the growing mode
is constant in time at leading order in the spatial gradient expansion. 

As shown in \cite{Takamizu:2010xy}, 
the linear solution valid up to $\calO(\epsilon^2)$ can be 
obtained as
\begin{eqnarray}
{\cal R}_{c,\bm{k}}^{\rm Lin}(\tau)
&=&\Bigl[\tilde{\alpha}^{\rm Lin}_{\bm{k}}+
(1-\tilde{\alpha}^{\rm Lin}_{\bm{k}}
){\tilde{D}(\tau)\over \tilde{D}_*}\nonumber\\
&&-\left({\tilde{F}_* \over \tilde{D}_*}\tilde{D}(\tau)+
\tilde{F}(\tau)\right)k^2\Bigr]U^{(0)}_{\bm{k}}\,,
\label{sol: linear Rc}
\end{eqnarray}
where the integrals $\tilde{D}(\tau)$ and 
$\tilde{F}(\tau)$ have been given as 
\begin{eqnarray}
\tilde{D}(\tau) & = & 3{\cal H}(\tau_*) 
  \int_{\tau_*}^\eta d\tau' {z^2(\tau_*) \over  z^2 (\tau')},
  \nonumber\\
 \tilde{F}(\tau) & = &
   \int_{\tau_*}^\tau \frac{d\tau'}{z^2(\tau')}
  \int_{\tau_*}^{\tau'}z^2(\tau'')c_s^2(\tau'')d\tau''.
\label{int-tilDF}
\end{eqnarray}
Here $\tilde{D}_*=\tilde{D}(\tau_*), \tilde{F}_*=\tilde{F}
(\tau_*)$, $\tau_*$ and $\cal H$ denote an initial time of 
gradient expansion and the conformal Hubble parameter 
${\cal H}=d \ln a/d\tau$, respectively.
The integrals in (\ref{int-tilDF}) represent a decaying and growing mode solution, respectively.

Note that ${\cal R}^{\rm Lin}_{c,_{\bm{k}}}(\tau_*)=U^{(0)}_{\bm{k}}$ that is 
just a constant solution, 
while ${\cal R}^{\rm Lin}_{c,_{\bm{k}}}(0)=\tilde{\alpha}^{\rm Lin}_{\bm{k}}U^{(0)}_{\bm{k}}$.
Thus if the factor $|\tilde{\alpha}^{\rm Lin}_{\bm{k}}|$ is large, it represents 
an enhancement of the curvature perturbation on superhorizon scales
due the $\calO(\epsilon^2)$ effect.

%%%%%%%%%%%%%%%%%%%%%%%%%%%%%%%%%%%%%%%%%%%%%%%%%%%%%%

Here it is useful to consider an explicit
expression for $\tilde{\alpha}^{\rm Lin}_{\bm{k}}$ in terms of 
${\cal R}^{\rm Lin}_{c,{\bm{k}}}$ and its derivative at $\tau=\tau_*$.
The result is 
\begin{eqnarray}
\tilde{\alpha}^{\rm Lin}_{\bm{k}}=1+ {\tilde{D}_*\over 3 {\cal H}_*}
\frac{{{\cal R}^{\rm Lin}_{c,_{\bm{k}}}}'(\tau_*)}{{\cal R}^{\rm Lin}_{c,_{\bm{k}}}(\tau_*)}
-k^2 \tilde{F}_*+\calO(k^4).
\end{eqnarray}

In order to relate our calculation with the standard formula for
the curvature perturbation in linear theory, we introduce 
$\tau_k$ (or $t_k$) which denotes the time at which the comoving wavenumber
has crossed the Hubble horizon,
\begin{equation}
\tau_k=-{r\over k}\,;\quad 0<r\ll 1 \,.
\end{equation}
The power spectrum at the horizon crossing time is given by 
%============< EQUATION >==============%
%
\begin{align}
 &\langle {\cal R}^{\rm Lin}_{c,\bm{k}}(\tau_k)
  {\cal R}^{\rm Lin}_{c,\bm{k}'}(\tau_{k'})
  \rangle
  = (2\pi)^3P_{{\cal R}}(k)\delta^3(\bm{k}+\bm{k}'), \nonumber\\
  &P_{{\cal R}}(k) = 
  \left|{\cal R}^{\rm Lin}_{c,\bm{k}}(\tau_k)\right|^2\,.
  \label{eq: power spectrum for R}
\end{align}
%======================================%

By inverting ${\cal R}^{\rm Lin}_{c,\bm{k}}$ in terms of $U^{(0)}_{\bm{k}}$ as 
shown in \cite{Takamizu:2010xy}, we can show the final value
of the linear curvature perturbation as 
\begin{equation}
{\cal R}^{\rm Lin}_{c,\bm{k}}(0)=
 \tilde{\alpha}^{\rm Lin}_{\bm{k}}U^{(0)}_{\bm{k}} = 
  {\alpha}^{\rm Lin}_{\bm{k}}
  {\cal R}^{\rm Lin}_{c,\bm{k}}(\tau_k)
  + \calO(k^4)\,,
  \label{eqn:alphau}
\end{equation}
%======================================%
where
%============< EQUATION >==============%
%
\begin{equation}
 {\alpha}^{\rm Lin}_{\bm{k}}
  =
  1 + \alpha^{\cal R}{D}_k- k^2 {F}_k \,,
  \label{til-alpha-lin} 
\end{equation}
%======================================%
and 
\begin{eqnarray}
 \alpha^{\cal R} & = & 
  \frac{1}{3{\cal H}(\eta_k) }
  {{\cal R^{\rm Lin}}'_{c,\bm{k}}\over {\cal R}^{\rm Lin}_{c,\bm{k}}}\bigg|_{\tau=\tau_k},
  \nonumber\\
 {D}_k & = & 3{\cal H}(\tau_k) 
  \int_{\tau_k}^0d\tau' {z^2(\tau_k) \over  z^2 (\tau')},
  \nonumber\\
 {F}_k & = &
 \int_{\tau_k}^0\frac{d\tau'}{z^2(\tau')}
  \int_{\tau_k}^{\tau'}z^2(\tau'')c_s^2(\tau'')d\tau''.
\end{eqnarray}
The formula (\ref{eqn:alphau}) will 
be used in the next subsection.

The power spectrum at the final time is thus enhanced by the factor 
$|{\alpha}^{\rm Lin}_{\bm{k}}|^2$ as
%============< EQUATION >==============%
%
\begin{equation}
 \langle {\cal R}^{\rm Lin}_{c,\bm{k}}(0)
  {\cal R}^{\rm Lin}_{c,\bm{k}''}(0)
  \rangle
  = (2\pi)^3  |{\alpha}^{\rm Lin}_{\bm{k}}|^2
  P_{{\cal R}}(k)\delta^3(\bm{k}+\bm{k}')\,. 
  \label{eqn:powerspectrum}
\end{equation}
%%%%%%%%%%%%%%%%%%%%%%%%%%%%%%%%%%%%%%%%%%%%%%%%
\subsection{Nonlinear theory valid up to $\calO(\epsilon^2)$}
Using the linear solution of the curvature perturbation 
given by (\ref{sol: linear Rc}), here we can derive
the nonlinear solution by matching the two at $\tau=\tau_*$. 
The main purpose of the matching is to make it possible to analyze
superhorizon nonlinear evolution valid up to the second order in
gradient expansion, starting from a solution in the linear theory. In
particular, we would like to evaluate the bispectrum induced by the
superhorizon nonlinear evolution. For this purpose, we need to
have full control over terms up not only to $\calO(\epsilon^2)$ but also
to $\calO(\delta^2)$, where we suppose that the linear solution is of order
$\calO(\delta)$. Therefore, the matching condition at $\tau=\tau_*$ should
be of the form 
%============< EQUATION >==============%
%
\begin{eqnarray}
 {\cal R}^{\rm NL}_c(\tau_*)
  & = & {\cal R}^{\rm Lin}_c(\tau_*) + s_1(\tau_*)
  + \calO(\epsilon^4,\delta^3), \nonumber\\
 {{\cal R}^{\rm NL}_c}'(\tau_*)
  & = & {{\cal R}^{\rm Lin}_c}'(\tau_*)
  + s_2(\tau_*) + \calO(\epsilon^4,\delta^3)\,,
\end{eqnarray}
%======================================%
where $ s_{1}(\tau_*) = \calO(\delta^2)$ and $s_{2}(\tau_*) = \calO(\delta^2)$ 
are functions of $\tau_*$ and spatial coordinates. While the linear
solution ${\cal R}^{\rm Lin}_c(\tau)$ is considered as an input,
i.e., initial condition, the additional terms, $s_{1}(\tau_*)$ and
$s_2(\tau_*)$, are to be determined by the following condition. 
The terms of order $\calO(\delta^2)$ in 
     ${\cal R}^{\rm NL}_{c,\bm{k}}$ and 
     ${{\cal R}^{\rm NL}_{c,\bm{k}}}'$ should vanish at the horizon crossing when $\tau=\tau_k$. Note that $\tau_k<\tau_*$. 
In other words, $s_1(\tau_*)$ and $s_2(\tau_*)$ represent the
$\calO(\delta^2)$ part of ${\cal R}^{\rm NL}_c$ and 
${{\cal R}^{\rm NL}_c}'$, respectively, generated during the period between
the horizon crossing time and the matching time. 

We have to omit the explicit way 
to determine the terms $s_1$ and $s_2$ for want of space, that was 
shown in \cite{Takamizu:2010xy}. 
As a result, using the linear solution of the curvature perturbation 
given by (\ref{sol: linear Rc}) we have the nonlinear
comoving curvature perturbation at the final time $\tau=0$ (or
$t=\infty$) given by 
\begin{eqnarray}
&&{\cal R}_{c,\bm{k}}^{\rm NL}(0)=
{\cal R}^{\rm Lin}_{c,\bm{k}} (\tau_k) 
-(1-{\alpha}^{\rm Lin}_{\bm{k}})
{\cal R}^{\rm Lin}_{c,\bm{k}} (\tau_k) \nonumber\\
&&
-\frac{1}{4}F_k
  \tilde{R}^{(2)}[{\cal R}^{\rm Lin}_c(\tau_k)]
 +\calO(\epsilon^4, \delta^3)\,,
\label{sol: infty Rreal}
\end{eqnarray}
where 
\begin{align}
\tilde{R}^{(2)}[\ell^0]&\equiv -2(\delta^{ij}\partial_i \ell^0
\partial_j\ell^0-4 \ell^0\Delta \ell^0)\nonumber\\
&=4\Delta \ell^0
+R^{(2)}[\ell^0]+\calO((\ell^0)^3)\,.
\end{align}
The first term in (\ref{sol: infty Rreal}) 
corresponds to the result of the $\delta N$ formalism, that is 
a constant since we considered the system for a single scalar field, 
the second term is related to an enhancement on superhorizon scales
in linear theory, and the last term is the nonlinear effect which
may become important if $F_k$ is large. 

Here we can notice that 
in order to the final values of curvature 
perturbation both in linear (\ref{eqn:alphau}) and in nonlinear theory (\ref{sol: infty Rreal}), all one have to do is to estimate the same integrals shown in both theories as $D_K$ and $F_k$ in $\alpha^{\rm Lin}_{\bm{k}}$. 
The reason why is 
that the master equations (\ref{ieq1}) and (\ref{ieq2}) for both theories have 
the same structures of evolution equation as described before.

In this subsection, we calculate the bispectrum of our nonlinear 
curvature perturbation by assuming that 
${\cal R}^{\rm Lin}_{c,\bm{k}}(\tau_k)$ is a Gaussian 
random variable. We assume the leading order contribution to the bispectrum
comes from the terms second order in ${\cal R}^{\rm Lin}_{c,\bm{k}}(\tau_k)$. 
The final result (\ref{sol: infty Rreal}) can be reduced to 
\begin{eqnarray}
 &&\zeta_{\bm{k}}={\cal R}^{\rm NL}_{c,\bm{k}}(0) \nonumber\\
 && =  
  \alpha^{\rm Lin}_{\bm{k}}\ {\cal R}^{\rm Lin}_{c,\bm{k}}(\tau_k)
 + {F_k\over 2}\ 
 \Biggl\{
  \int\frac{d^3k'd^3k''}{(2\pi)^3}
    (4{k'}^2-\delta_{ij}{k'}^i{k''}^j)\nonumber\\
&&\times 
    {\cal R}^{\rm Lin}_{c,\bm{k}'}(\tau_{k'})
    {\cal R}^{\rm Lin}_{c,\bm{k}''}(\tau_{k''})
    \delta^3(-\bm{k}+\bm{k}'+\bm{k}'')
 \Biggr\} \nonumber\\
 && + \calO(\epsilon^4, \delta^3)\,.
  \label{eq: final-zeta-k}
\end{eqnarray}
%======================================%
By assuming the Gaussian statistics for 
${\cal R}^{\rm Lin}_{c,\bm{k}}(\tau_k)$, 
it is easy to calculate the power spectrum shown as (\ref{eqn:powerspectrum}) 
with (\ref{eq: power spectrum for R}) and the bispectrum of
primordial curvature perturbation: $\zeta$. 

The dimensionless bispectrum ${\cal B}_{\zeta}$ is expressed in terms of the 
Fourier transformation of the three point function as
\begin{eqnarray}
\left\langle{\zeta}_{{\bm k}_1}{\zeta}_{{\bm k}_2}{\zeta}_{{\bm k}_3}
\right\rangle_C
=(2\pi)^7 \delta^{3} ({\bf k}_1+{\bf k}_2+{\bf k}_3)
{\cal P}^2_\zeta {{\cal B}_{\zeta}\over k_1^2 k_2^2 k_3^2},
\end{eqnarray}
where $\langle\cdots\rangle_C$ means that it extracts out
only connected graphs. We use the dimensionless quantity ${\cal B}_\zeta$ 
to represent the amplitude of the bispectrum with the 
uncorrected power spectrum ${\cal P}_\zeta$, which has been defined by 
${\cal P}_\zeta = 
k^3 P_\zeta/2 \pi^2$. We can use a standard amplitude of dimensionless 
power spectrum as ${\cal P}_\zeta=\calO(10^{-9})$. 
With the help of (\ref{eq: final-zeta-k}), 
the three point correlation function of ${\zeta}$ is at leading order
calculated as
\begin{eqnarray}
&&\left\langle{\zeta}_{{\bm k}_1}{\zeta}_{{\bm k}_2}{\zeta}_{{\bm k}_3}
\right\rangle_C=(2\pi^3)
\Biggl[Re(\alpha^{{\rm Lin}*}_{k_1}\alpha^{\rm Lin}_{k_2})F_{k_3}\nonumber\\
&&\times 
\left\{ 2(k_1^2+k_2^2)-\delta_{ij}k_1^ik_2^j\right\}\delta^{(3)}
({\bf k}_1+{\bf k}_2+{\bf k}_3)\nonumber\\
&&\times 
|{\cal R}^{\rm Lin}_{c,\bm{k}_1}(\tau_{k_1})|^2
|{\cal R}^{\rm Lin}_{c,\bm{k}_2}(\tau_{k_2})|^2+ {\rm 2 terms} \Biggr]\,,
\end{eqnarray}
where $Re$ means taking a real part, a superscript star denotes a complex conjugate and `2 terms' means 
terms with cyclic and permutations among the three wavenumbers.
The power spectrum of ${\cal R}^{\rm Lin}_{c,\bm{k}}(\tau_k)$ is written
as (\ref{eq: power spectrum for R}). Then we have 
\begin{eqnarray}
&&{\cal B}_{\zeta}(k_1,k_2,k_3)  =  
\frac{1}{8 k_1k_2k_3}\Biggl[
Re(\alpha^{{\rm Lin}*}_{k_1}\alpha^{\rm Lin}_{k_2})F_{k_3}\nonumber\\
&&\times\left\{5(k_1^2+k_2^2)-k_3^2\right\}
k_3^3\,\triangle{\cal P}_{{\zeta}}(k_1)
\triangle{\cal P}_{{\zeta}}(k_2)\notag\\
&&+ {\rm 2 terms} \Biggr]
\,,
\label{eq: bispectrum-zeta}
\end{eqnarray}
where $\triangle{\cal P}_{\zeta}$ denotes the modulation factor of power spectrum, that is 
a ratio of a corrected power spectrum to uncorrected one: 
\begin{eqnarray}
\triangle{\cal P}_{\zeta}(k)= 
{k^3\over 2\pi^2{\cal P}_{\zeta} }
\left|{\cal R}^{\rm Lin}_{c,\bm{k}}(\tau_k)\right|^2\,.
\end{eqnarray}
%%%%%%%%%%%%%%%%%%%%%%%%%%%%%%%%%%%%%%%%%%%%%%%%%
\section{Application--varying sound speed--}
\begin{figure}[tbp]
\begin{center}
\includegraphics[width=7cm]{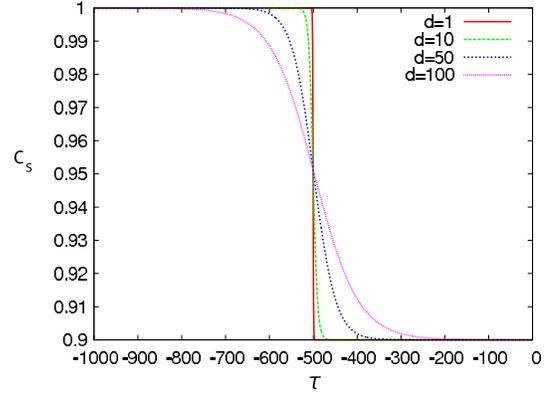} 
\end{center}
\vspace{-0.5cm}
\caption{We plot varying sound speed as taking $c_{s1}=1, c_{s2}=0.9$ for various values of varying width $d$. We set $\tau_0=-500$. The horizontal axis is taking the conformal time.}
\label{cs}
\end{figure}

We consider the case of varying sound speed as one application of beyond $\delta N$-formalism. As a simple example, 
we have assumed that the background evolution satisfies the slow-roll conditions throughout this paper, that is
\begin{equation}
\eta_1=-{\dot H\over H^2}\ll 1
\ {\rm and}\  \eta_2={\dot \eta_1\over H \eta_1}\ll 1\,,
\end{equation}
where a dot denotes a derivatives with respect to the physical time $t$. 
We compute the curvature perturbation for a model such that 
time variation of the sound speed is described by the following function as 
\begin{eqnarray}
c_s^2=c_{s1}^2+(c_{s2}^2-c_{s1}^2){\tanh[(\tau-\tau_0)/d]+1\over 2}\,,
\end{eqnarray}
where $c_{s1}, c_{s2}, \tau_0$ and $d$ are parameters and the sound speed 
changes from $c_{s1}$ to $c_{s2}$ with a varying duration characterised by 
$\tau_0$ and $d$. We can introduce a new parameter $T=c_{s1}/c_{s2}$, which 
represents the ratio of sound speed before and after the transition. 
If we take the width of duration very small $d\lsim 1$, the model results in the previous study of sudden varying sound speeds as in 
\cite{Nakashima:2010sa}. Throughout this paper, 
we set $\tau_0=-500$, but the results do not 
depend on specifying the choice of the parameter. 
We plot the evolution of sound speed for example in Fig.\ref{cs} where we set $T=0.9$ and $d=1,10,50,100$. 
%%%%%%%%%%%%%%%%%%%%%%%%%%%%%%%%%%%%%%%%
%%%%%%%%%%%%%%%%%%%%%%%
%%%%%%%%%%%%%%%%%%%%%%%%%%%%%%
\begin{figure*}[t]
\begin{center}
\begin{tabular}{ll}
%\vspace{-1cm}
\includegraphics[width=7cm]{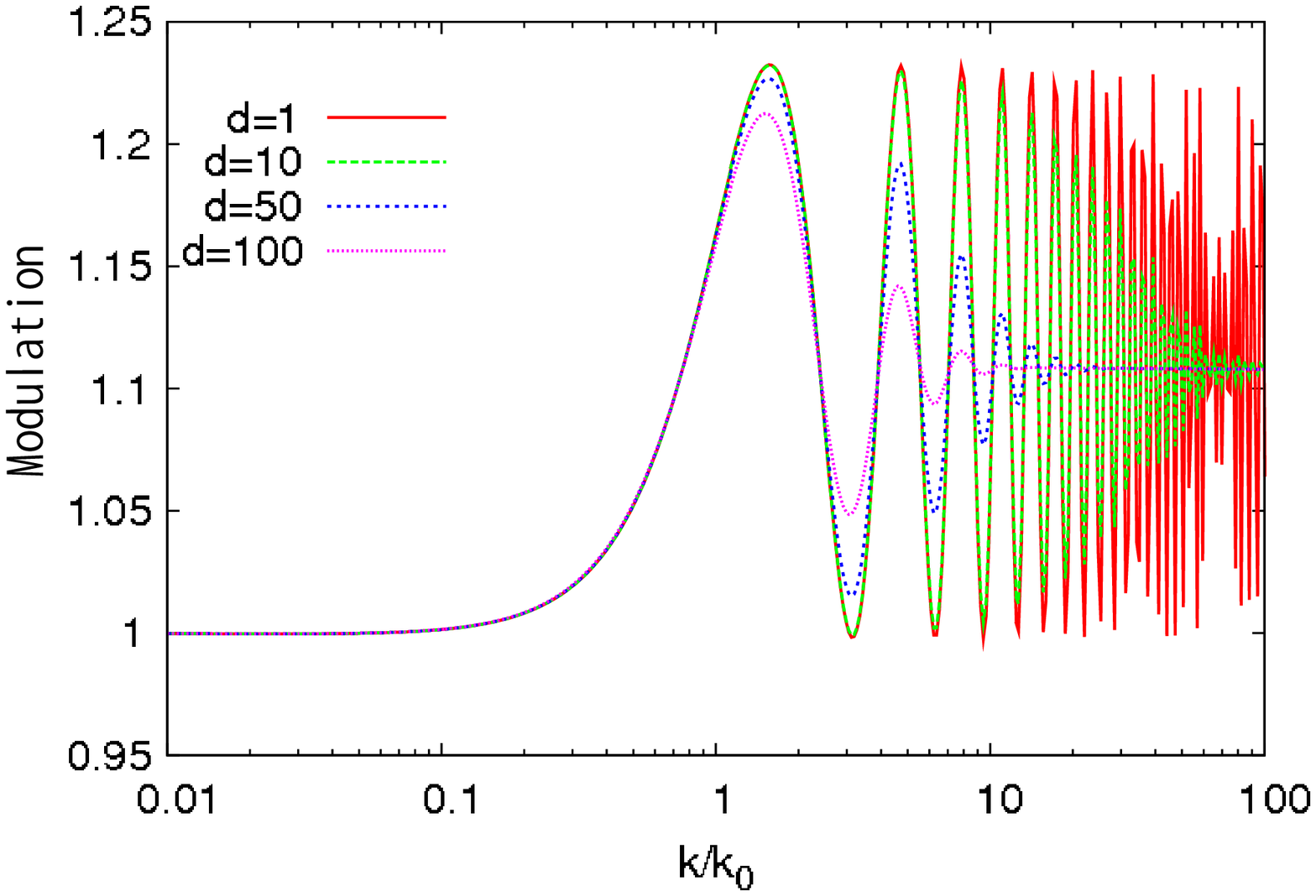} & 
\hspace{1cm}
\includegraphics[width=7cm]{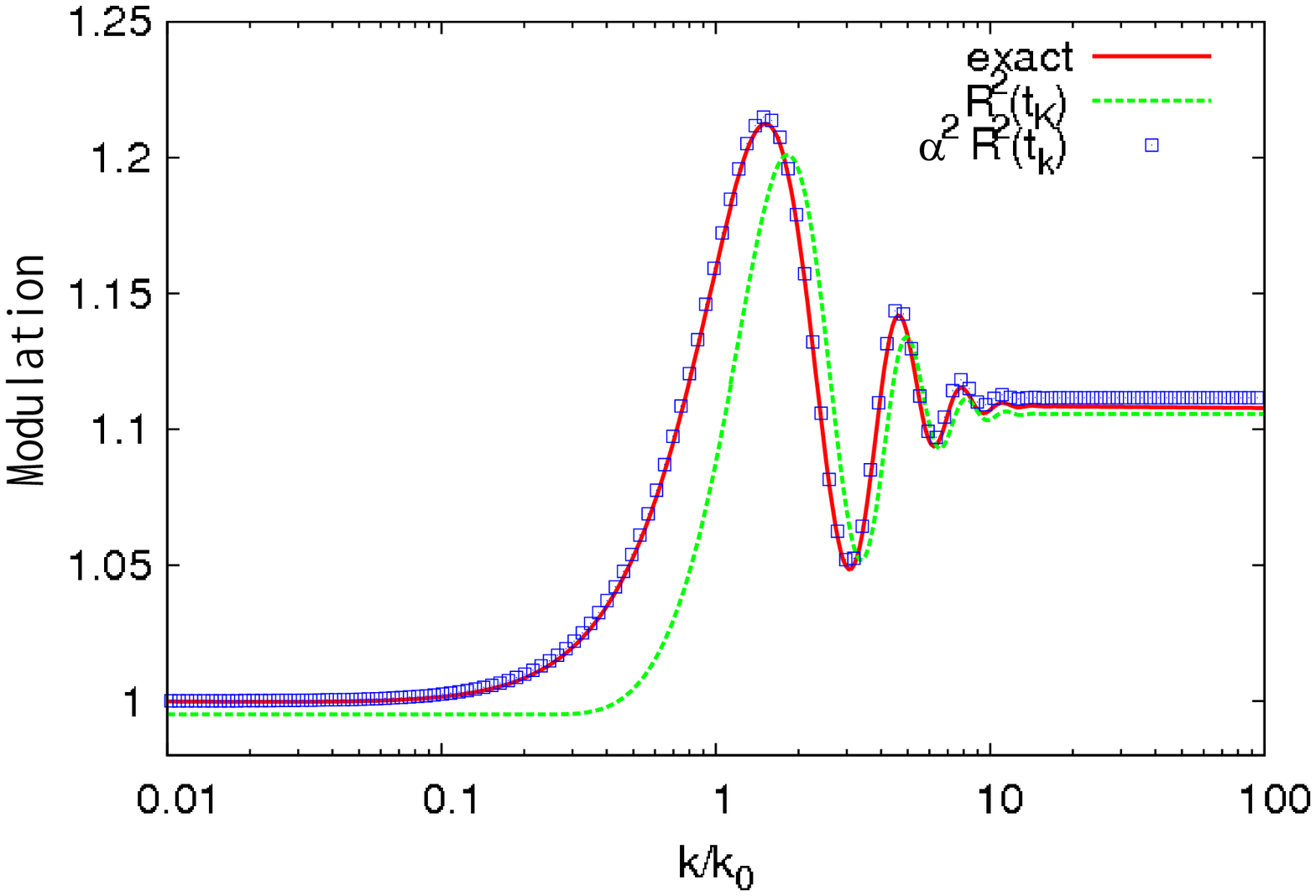} 
\end{tabular}
\vspace{-0.5cm}
\end{center}
\caption{(Left) We plot the modulation factor of final power spectrum for various values of $d$ with 
setting $T=0.9$.  
We can see the featured bumps at $k=k_0$ with oscillation. As $d$ taking smaller value, 
the oscillations are more intensive. (Right) Comparison the long-wavelength approximation (\ref{eqn:powerspectrum}) 
with the numerical exact solution of (\ref{basic_uk}). We set $T=0.9$ and $d=100$. 
The enhancement from the amplitude at the horizon crossing time ${\cal R}^2
(\tau_k)$(green line), which is described by $(\alpha^{\rm Lin}_{\bm k})^2$ of (\ref{til-alpha-lin}), occurs at superhorizon scales $k/k_0<1$.
}
\label{Rk}
\end{figure*}
%%%%%%%%%%%%%%%%%%%%%%%%%%%%%%%%%%%%%%%%%
\begin{figure*}[t]
\begin{center}
\begin{tabular}{ll}
\includegraphics[width=7cm]{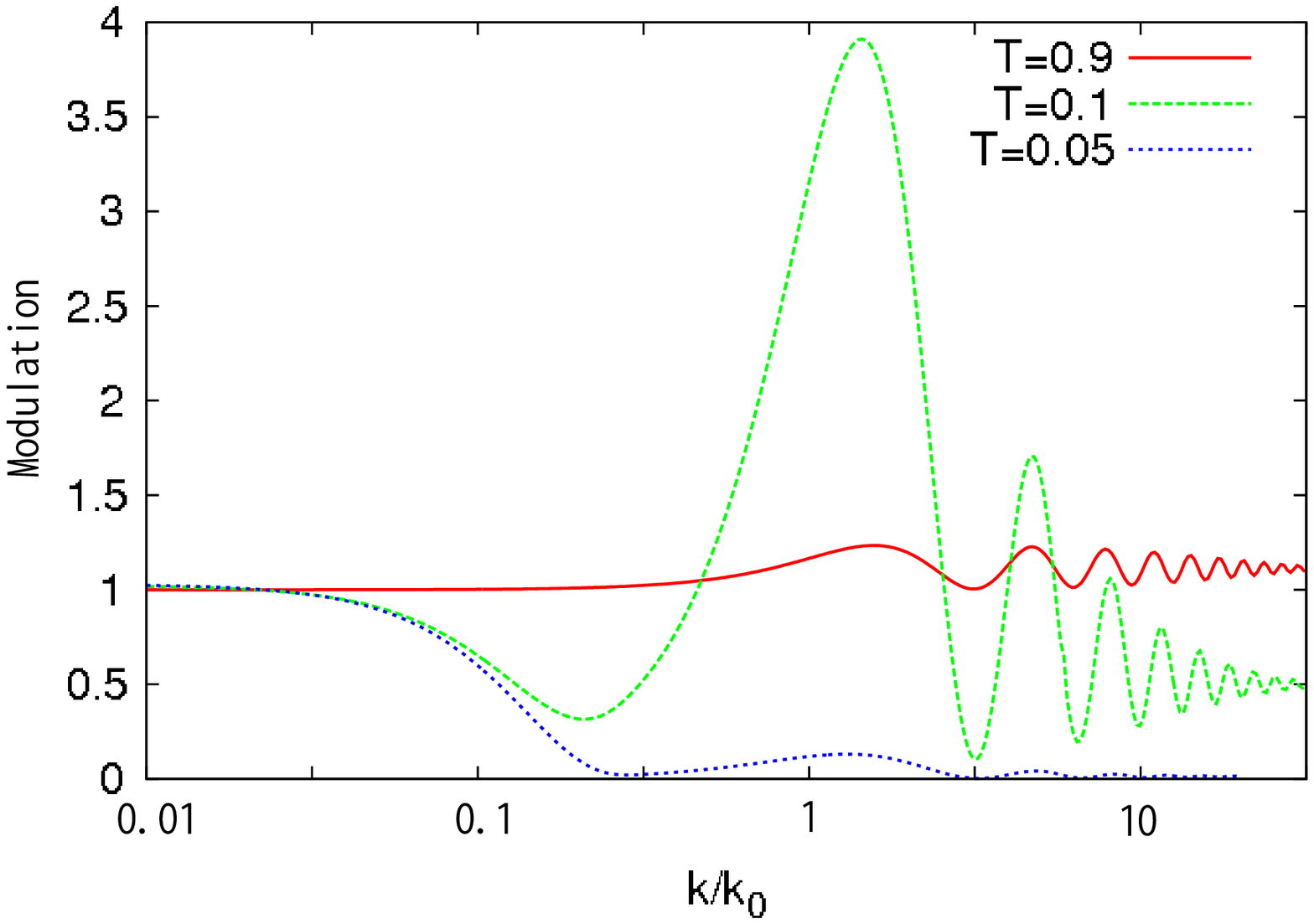} & 
\hspace{1cm}
\includegraphics[width=7cm]{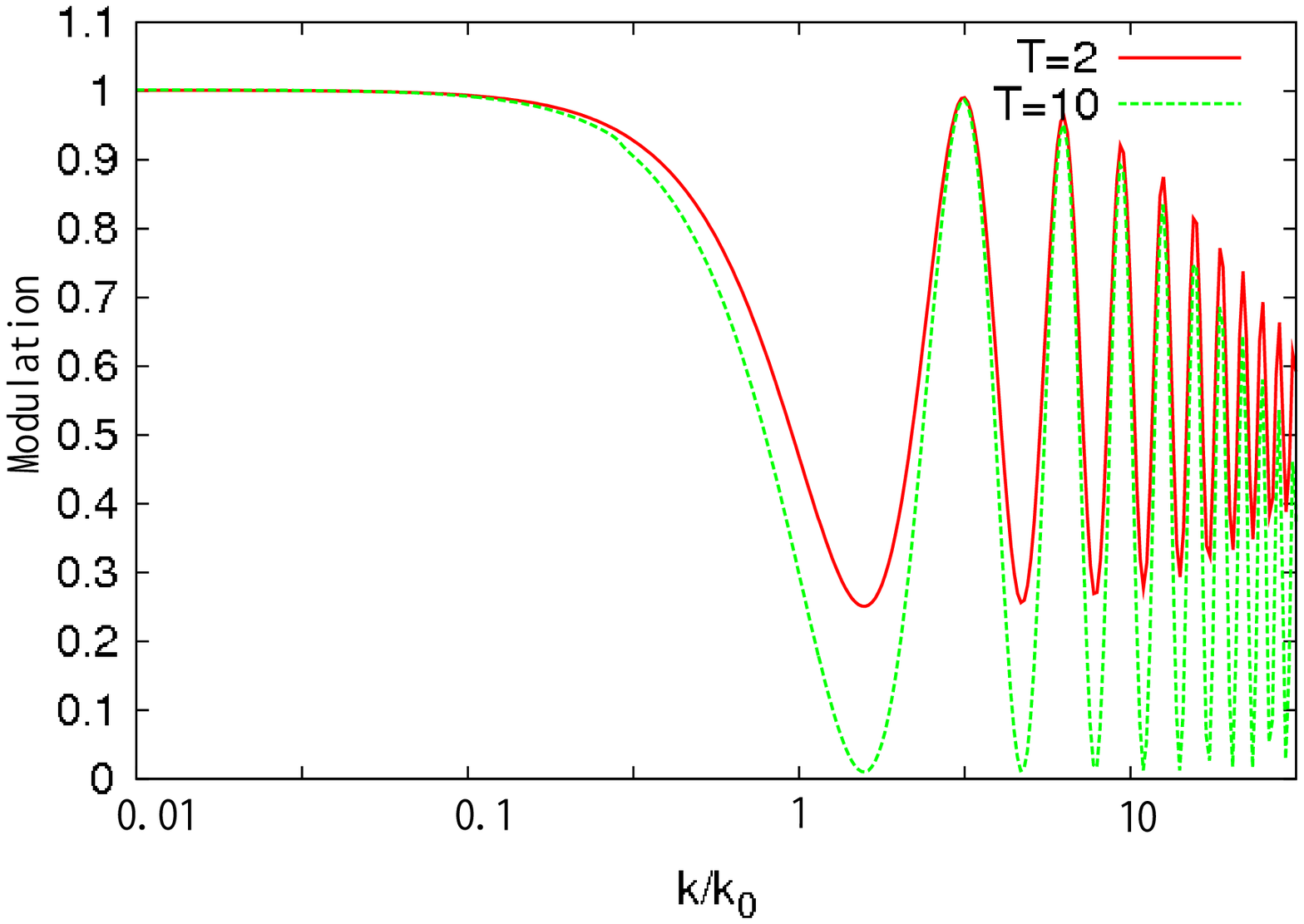}
\end{tabular}
\end{center}
\vspace{-0.5cm}
\caption{We plot the modulation of final power spectrums for various values of $T<1$(left) and 
$T>1$(right) with setting $d=20$. All cases in left(right) panel correspond to the situations of decreasing(increasing) sound speeds. }
\label{Rk-T}
\end{figure*}
%%%%%%%%%%%%%%%%%%%%%%%%%%%%%%%%%%%%%%
\begin{figure*}[t]
\vspace{-1cm}
\begin{center}
\begin{tabular}{ll}
\includegraphics[width=6.5cm]{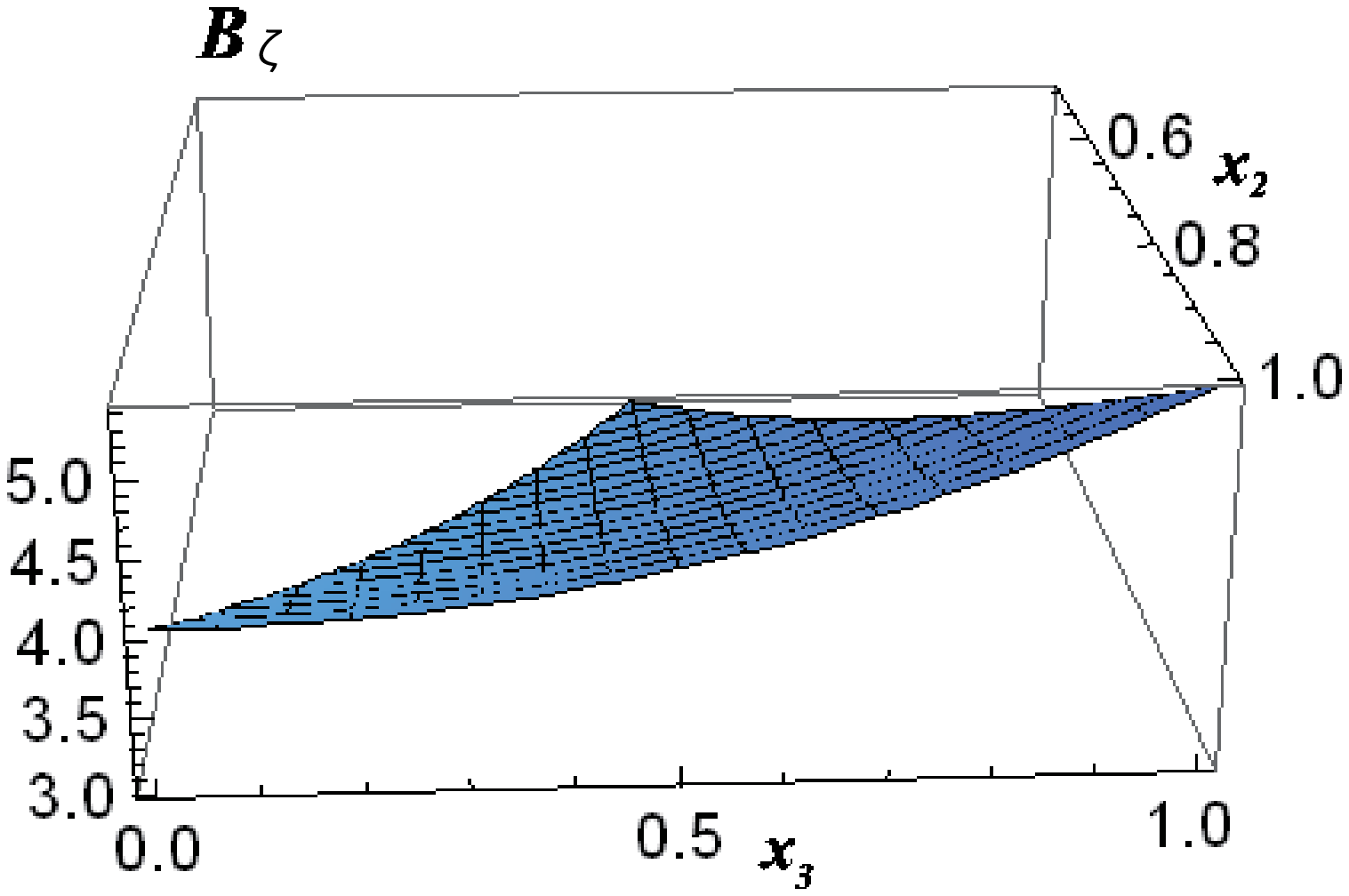} & 
\hspace{1cm}
\includegraphics[width=6.5cm]{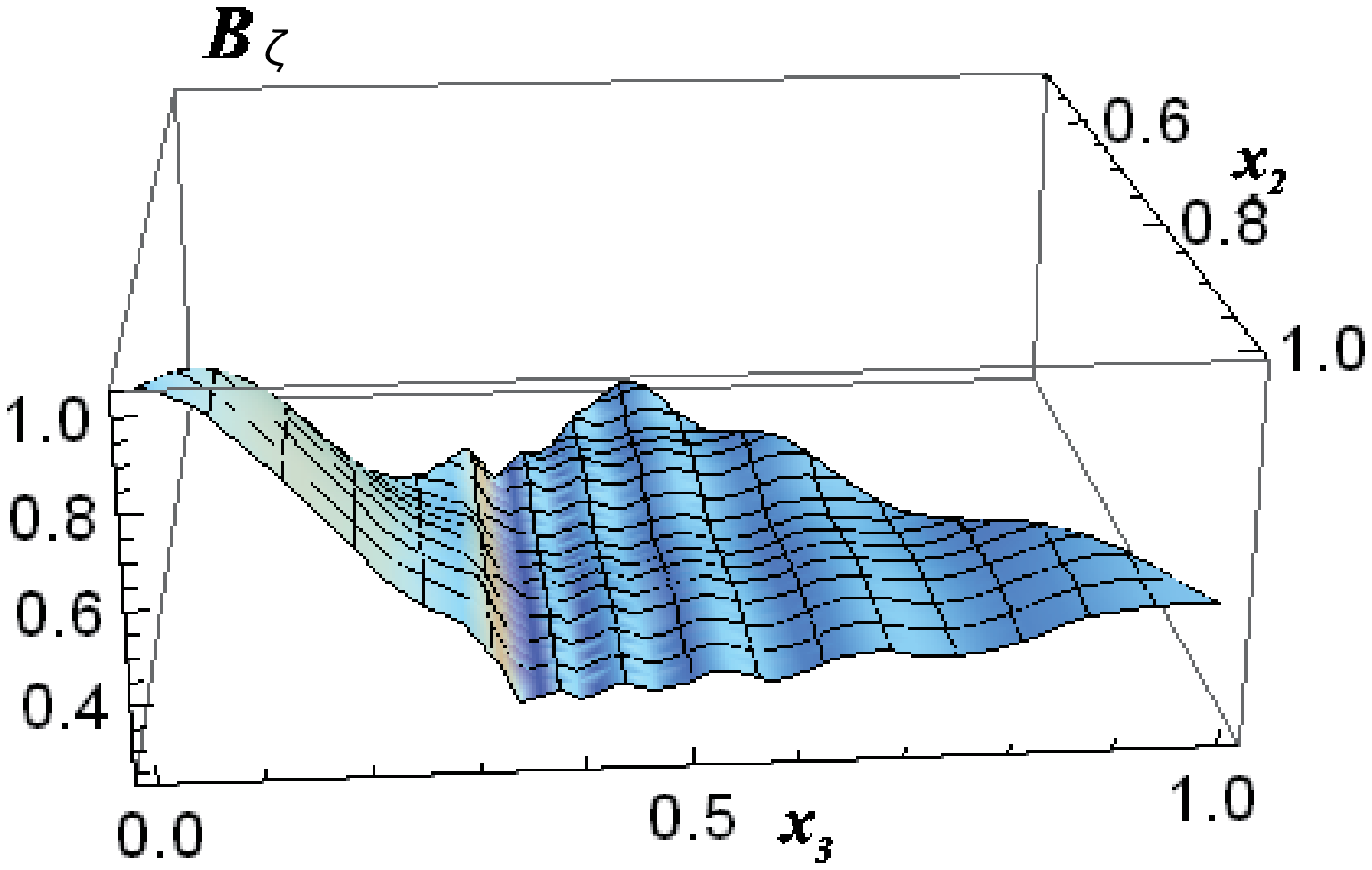}
\end{tabular}
\end{center}
\vspace{-1cm}
\caption{We plot dimensionless bispectrum as a function of $x_2=k_2/k_1$ and $x_3=k_3/k_1$. We set $T=0.5, d=20$ with parameters: $k_1/k_0=0.1$(left) and 
$k_1/k_0=1$(right). The bispectrum have a peak 
at the equilateral(local) shape in left(right) panel. }
\label{Bis}
\end{figure*}
%%%%%%%%%%%%%%%%%%%%%%%%%%%%%%%%%%
%%%%%%%%%%%%%%%%%%%%%%%%%%%%%%%
\subsection{Power Spectrum}
The basic equation in the linear theory 
for primordial curvature perturbation $\zeta$ is 
written in terms of $v=\zeta z$. 
The basic equation of motion for Fourier modes are given by
\begin{equation}
v_{k}^{\prime\prime}+\left(c_{s}^{2}k^{2}-\frac{z^{\prime\prime}}{z}\right)v_{k}=0\,.
\label{eq: basic eq-v}
\end{equation}
We introduce a variable $u$, which is related to $v$ as 
\begin{equation}
-c_{s}k^{2}u_{k}=z\left(\frac{v_{k}}{z}\right)^{\prime},
\ \ c_{s}v_{k}=\theta\left(\frac{u_{k}}{\theta}\right)^{\prime}\,,
\label{eq: relation-u-v}
\end{equation}
where we have defined $
\theta\equiv{1}/{(c_{s}z)}$. 
The basic equation of motion (\ref{eq: basic eq-v}) in terms of 
the Fourier modes $u_k$ is obtained as 
\begin{equation}
u_{k}^{\prime\prime}+\left(c_{s}^{2}k^{2}-\frac{\theta^{\prime\prime}}{\theta}\right)u_{k}=0\,.
\label{eq: basic eq-u}
\end{equation}
Note that the term $(c_{s}^{\prime}/c_{s})$ does not 
exist in $\theta''/\theta$ since the variable $\theta$ does not 
depend on $c_s$ from (\ref{zdef}) as $\theta=1/(a\sqrt{2 \eta_1})$. 
We have to solve this equation 
under the background evolution. 
The term $\theta''/\theta$ is rewritten in terms of slow-roll parameters as 
\begin{align}
\frac{\theta^{\prime\prime}}{\theta}=\frac{1}{\tau^{2}}\left(\frac{\eta_2}{2}-\eta_1\right)\,,
\end{align}
where we have used slow-roll approximation that is, 
$|\eta_1|,|\eta_2|\ll 1$ and taking their linear limits, and used 
a useful equation 
\begin{equation}
aH=-\frac{1}{\tau(1-\eta_1)}\,.
\end{equation}
Therefore, we can obtain the basic equation 
\begin{equation}
u_{k}^{\prime\prime}+\left(c_{s}^{2}k^{2}-\frac{\nu^{2}-\frac{1}{4}}{\tau^{2}}\right)u_{k}=0\,,
\label{basic_uk}
\end{equation}
where we have defined 
\begin{equation}
\nu^{2} = \frac{\eta_2}{2}-\eta_1+\frac{1}{4}\,,
\end{equation}
and approximate it as 
\begin{equation}
\nu =\sqrt{\frac{\eta_2}{2}-\eta_1+\frac{1}{4}} \approx 
\frac{1}{2}+\frac{\eta_2}{2}-\eta_1\,.
\end{equation}
In the regime when $\tau < \tau_{0}$, setting $c_{s}=c_{s1}$ leads to 
the equation of motion 
\begin{equation}
u_{k}^{\prime\prime}+\left(c_{s1}^{2}k^{2}-\frac{\nu^{2}-\frac{1}{4}}{\tau^{2}}\right)u_{k}=0\,,
\end{equation}
and its solution is obtained by 
\begin{equation}
u_{k1}=\sqrt{-kc_{s1}\tau}\left[c_{1}H_{\nu}^{(1)}(-kc_{s1}\tau)+c_{2}H_{\nu}^{(2)}(-kc_{s1}\tau)\right]\,,
\end{equation}
where $H_{\nu}^{(1),(2)}(\tau)$ denote the Hankel function and 
$c_{1,2}$ are arbitrary constants, which have to be 
determined by initial conditions at the time $\tau\to -\infty$. 
We choose the adiabatic vacuum at the initial time in terms of 
$v_k$ as $v_{k} \to e^{-ikc_{s1}\tau}/{\sqrt{2kc_{s1}}}$.
Hence it leads to the choice of $c_1,c_2$ as
\begin{equation}
c_{1}=\frac{i}{2k^{3/2}}\sqrt{\frac{\pi}{c_{s1}}}\exp\left(\frac{2\nu+1}{4}\pi i\right), \ \ \ c_{2} =0\,.
\end{equation}
We solve the basic equation (\ref{basic_uk}) numerically with the above initial condition. This solution can show us the evolution from subhorizon scale to superhorizon scale. On the other hand, we can calculate the enhance factor 
$|\alpha^{\rm Lin}_{\bm{k}}|$ by estimating the equation (\ref{eqn:powerspectrum}) obtained under the long-wavelength approximation. Then we can compare it with the above numerical exact solution. In order to compare them, we have to 
estimate $\zeta$ from the numerical solution of $u$ by using the relation 
$\zeta=v/z=\theta^2(u/\theta)'$. 

First we will show the exact solution by 
using numerical solving in the left panel of Fig.\ref{Rk} for various values of $d$. We plot the modulation factors of power spectrum with $k/k_0$, where $k_0$ is the wavenumber 
corresponding to a transition time $\tau_0$. 
It shows some feature like bump at $k=k_0$ 
with oscillation. As $d$ taking smaller value, 
the oscillations are more intensive and they do not converge 
for $d<\calO(1)$. 
 The result of $d=1$ is consistent with 
the previous result of Ref.\cite{Nakashima:2010sa} for studying the case of 
a sudden varying sound speed. 

The right panel of Fig.\ref{Rk} shows 
the comparison such exact solution with solution obtained by using 
the long-wavelength approximation as (\ref{eqn:powerspectrum}). 
It tells us that 
the approximation is very good for fitting the exact solution. Especially, we 
can see that the approximation is good for not only the 
superhorizon regime $k/k_0<1$, but also the subhorizon regime $k/k_0>1$. 
The enhancement from the amplitude at the horizon crossing time, which is described by $(\alpha^{\rm Lin}_{\bm k})^2$ of (\ref{til-alpha-lin}), occurs at superhorizon scales $k/k_0<1$.

Next, we will examine how the modulation factors depend on 
different variables. 
In Fig.\ref{Rk-T}, we plot the final power spectrums for the case of 
decreasing sound speed $T<1$ and increasing one $T>1$, respectively. 
From the observation of WMAP, the parameter $|T-1|\gsim 0.1$ is strongly 
constrained, therefore the modulation appearing for $T=0.9$ is at most in order to make features in power spectrum (see \cite{Nakashima:2010sa} the details). 
%%%%%%%%%%%%%%%%%%%%%%%%%%%%%%%%%%%%%%%%
\subsection{Bispectrum}
In order to compare with observations, 
we can define a $k$-dependent nonlinear parameter by dividing the 
dimenionless bispectrum by a square of the corrected 
power spectrum at the final time: 
$f_{NL}(k_1,k_2,k_3)$ as
\begin{align}
&f_{NL}\equiv
{10\over 3}{k_1 k_2 k_3{\cal B}_\zeta \over 
{|\alpha^{\rm Lin}_{\bm k_1}
\alpha^{\rm Lin}_{\bm k_2}|^2
\triangle{\cal P}_\zeta(k_1)\triangle{\cal P}_\zeta}(k_2)k_3^3+
{\rm 2 terms}}
\notag\\
&=\frac{5}{12}\Biggl[|\alpha^{\rm Lin}_{\bm k_1}
\alpha^{\rm Lin}_{\bm k_2}|^2 k_3^3+
{\rm 2 terms}\Biggr]^{-1}\notag\\
&\times\Biggl[
Re(\alpha^*_{k_1}\alpha_{k_2})F_{k_3}\left\{5(k_1^2+k_2^2)-k_3^2\right\}
k_3^3+ {\rm 2 terms} \Biggr]\,.
\end{align}
%%%%%%%%%%%%%%%%%%%%%%%%%%%%%%%%%%%%%
\begin{figure}[bp]
\begin{center}
\includegraphics[width=7cm]{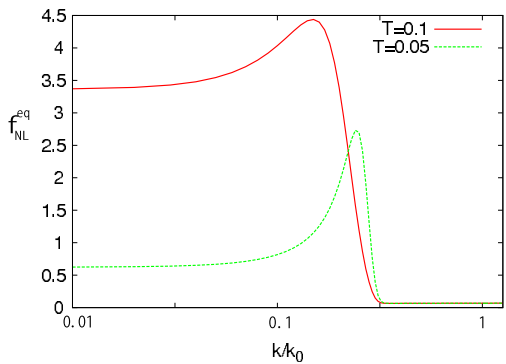} 
\end{center}
\vspace{-0.5cm}
\caption{We plot the equilateral bispectrum  
for various values of $T<1$ with $d=20$. They show the featured bispectra at 
$k/k_0\simeq 0.1$. }
\label{feq}
\end{figure}
%%%%%%%%%%%%%%%%%%%%%%%%%%%%%%%%%
\begin{figure*}[t]
\begin{center}
\begin{tabular}{ll}
\includegraphics[width=7cm]{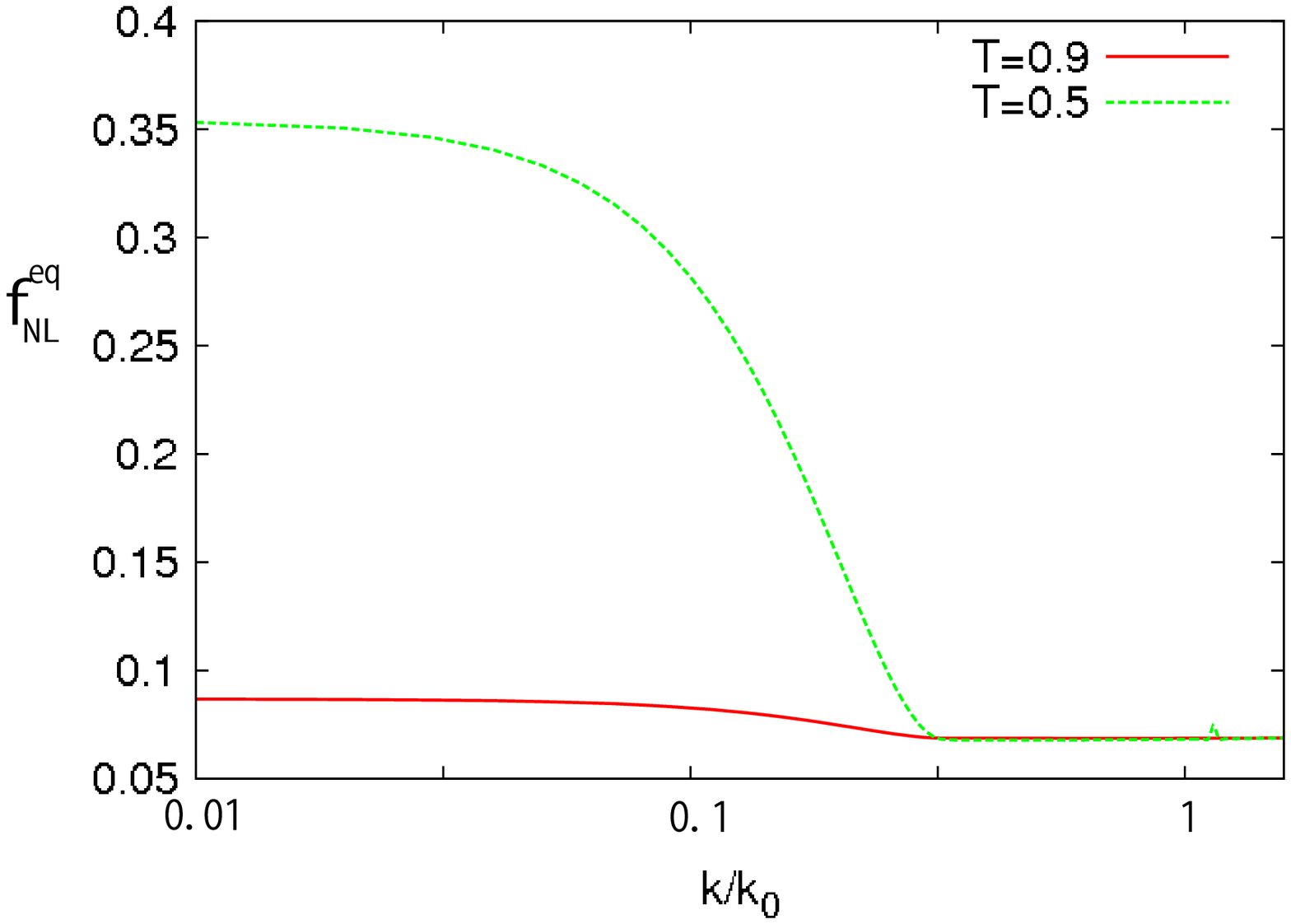} & 
\hspace{1cm}
\includegraphics[width=7cm]{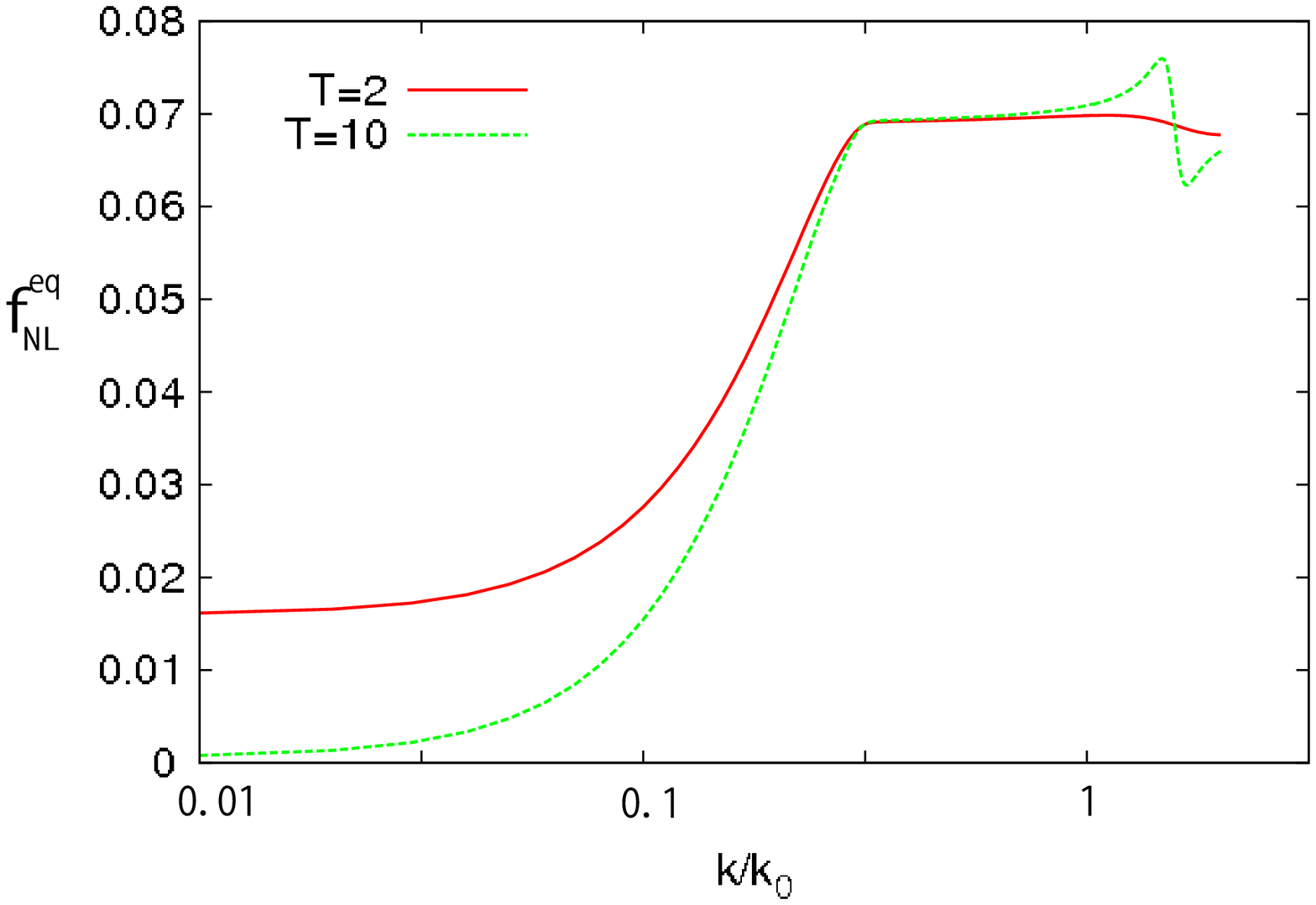} 
\end{tabular}
\end{center}
\vspace{-0.5cm}
\caption{The equilateral bispectrum  for $T<1$(left) and $T>1$(right) with $d=20$. Even though we can see small feature at subhorizon scale $k/k_0> 1$ for large value of $T=10$, the plots generally show no detectable feature.}
\label{feq-T}
\end{figure*}
%%%%%%%%%%%%%%%%%%%%%%%%%%%%%%%%%%%%%%%%%%%%%%%
Next we will plot $k$-dependent nonlinear parameter $f_{NL}(k_1,k_2,k_3)$. 
We plot a dimensionless bispectrum as a function of $x_2\equiv k_2/k_1$ and 
$x_3\equiv k_3/k_1$ with the free parameter $k_1/k_0$ in Fig.\ref{Bis}. 
In the figure, we take $k_1/k_0=0.1$ and $k_1/k_0=1$, respectively for both same settings of $T=0.5, d=20$. Here we notice that 
our expansion technique is valid for $k_1/k_0\leq 1$, since we can 
predict the evolution only when the transition happens after the horizon 
crossing, $t_k<t_0$.  
As shown in Fig.\ref{Bis}, 
for small value of $k_1/k_0<1$, the bispectrum has a 
peak at an equilateral shape; $k_1=k_2=k_3$. On the other hand, 
for $k_1/k_0=1$, it has a peak at a local(squeezed) shape; $k_3=0, k_1=k_2$. 

When we focus on bispectrum for superhorizon scales, i.e., taking $k_1/k_0<\calO(1)$, all bispectra have 
peaks at equilateral shape affected by the effect of finite changing 
duration time $d$, otherwise the delta approximation; $d\ll 1$ also show 
the local type of bispectra, which have been seen in the previous paper \cite{Park:2012rh}. Our 
results do not depend on specifying the choice of parameter $d$, 
only when we consider  a finite duration time; $d>\calO(1)$. 

Therefore, we will plot the equilateral bispectrum 
$f_{NL}^{eq}=f_{NL} (k_1=k_2=k_3)$ in Fig.\ref{feq}, and Fig.\ref{feq-T} for various values of $T$ with $d=20$. We can see the featured bispectrum in Fig.\ref{feq} where we take 
small values of $T$ as $T=0.1$ and $T=0.05$, pointing 
$f^{eq}_{NL}=\calO(5)$ 
within the recent constraint. 
On the other hand,  the cases for other values of $T$ show 
no such feature in Fig.\ref{feq-T}, where the equilateral bispectrum increases(decreases) towards super(sub)horizon scale as seen in the left(right) panel. 
We can also see small feature at subhorizon scale $k/k_0\gsim 1$ for large value of $T=10$, however this value of amplitude of the feature is too 
small to be detectable. 

%%%%%%%%%%%%%%%%%%%%%%%%%%%%%%%%%%%%%%%%%%%%%%%
\section{Concluding remarks}
We focus on the evolution 
of curvature perturbation on superhorizon scales by adopting 
the spatial gradient expansion. We  have reviewed such approximation in 
both linear and nonlinear theory, which is called the 
{\it beyond} $\delta N$-formalism as the next-leading order in the expansion. 
In our formalism \cite{Takamizu:2010xy,Takamizu:2013gy}, we can deal with the 
time evolution in contrast to $\delta N$-formalism, where 
curvature perturbations remain just constant, and nonlinear curvature perturbation follows the simple master equation whose form is similar as one in linear theory. 

As seen in (\ref{ieq1}) and (\ref{ieq2}), the evolution equation for 
curvature perturbation in both theories take similar structures, therefore in order to estimate the power spectrum (\ref{eqn:powerspectrum}) 
and bispectrum (\ref{eq: bispectrum-zeta}) in the approximation, 
all have to do is to calculating the same integrals as $D_K$ and $F_k$ in the 
enhancement factor: 
$\alpha^{\rm Lin}_{\bm{k}}$ shown in (\ref{til-alpha-lin}). It is 
easy to estimate non-Gaussianity, in contrast with 
the usual in-in formalism \cite{Maldacena:2002vr}, where 
a numerical calculation of the correlated function would be 
too difficult to solve (see also \cite{Chen:2006xjb} for the numerical method), if one consider a complicated situation need to solve numerically. Beyond $\delta N$-formalism 
takes an advantage to calculating the correlated features for power spectrum 
and bispectrum since the calculation is basically as same as solving the power spectrum. 

As one application of our formalism for a single scalar field, we investigate 
the case of varying sound speed. The previous 
studies \cite{Khoury:2008wj,Nakashima:2010sa,Park:2012rh,Emery:2012sm,Achucarro:2012fd,Bartolo:2013exa} 
have done for a sudden changing, but in this paper we studied 
a changing with a finite duration time need to solve numerically (see 
\cite{Achucarro:2012fd}, which also studied such mild transit in the speed of sound and also \cite{Achucarro:2013cva} 
for recent analysis by using PLANCK data). The main 
purpose of this paper is to analyze whether the features can 
appear in the bispectrum, in particular of equilateral shape 
by using our nonlinear perturbation theory. 
The case is more suitable to calculate by using our formalism than by using the in-in formalism. 
We discuss local features of primordial power and bispectrum generated by the effect of varying sound speed. 
As shown in \cite{Takamizu:2010xy} by using a similar way, we have also 
investigated one application of the beyond $\delta N$ for 
analyzing the featured bispectrum affected by 
a sharp change in the inflaton's potential slope. 

As shown in Fig.\ref{feq}, 
we can see a local feature like a bump at $k/k_0=\calO(0.1)$ for a 
small value of $T=c_{s1}/c_{s2}<\calO(0.1)$ in the equilateral bispectrum, 
which has a peak value of non-Gaussianity; $f_{NL}^{eq}=\calO(10)$ at most, 
consistent within the recent observational constraint by PLANCK. 
However, such parameters also lead to the features in the power spectrum, 
which are excluded from the observations since the 
current CMB experiment gives a strong constraint, 
which is sensitive to $|T-1|\gsim 0.1$ by 
CMB temperature power spectrum (see \cite{Nakashima:2010sa}). 

This study is one toy model as 
a first step to investigate more realistic situation, that is for example, 
including a background evolution, extending to multi-field system, etc. 
We plan 
to work on this and hope 
to discuss them in the future. 

%%%%%%%%%%%%%%%%%%%%%%%%%%%%
\section{Acknowledgments}
We thank Ryo Saito for useful discussion and comments. The work is 
supported by a Grant-in-Aid through JSPS No.24-2236. 
%%%%%%%%%%%%%%%%%%%%%%%%%%%%%%%%%%%%%%%%%%

\end{document}